\documentclass[letterpaper]{article}
\usepackage{spconf}  
\usepackage{amsmath,amssymb,amsfonts}
\usepackage{algorithmic}
\usepackage{graphicx}
\usepackage{textcomp}
\usepackage{xcolor}
\usepackage{comment}
\usepackage{booktabs}
\usepackage{multirow}
\usepackage{adjustbox} 
\usepackage{enumitem}
\usepackage[backend=biber,style=ieee]{biblatex}

\defbibheading{bibliography}[\refname]{}
\usepackage{balance}
\def\BibTeX{{\rm B\kern-.05em{\sc i\kern-.025em b}\kern-.08em
    T\kern-.1667em\lower.7ex\hbox{E}\kern-.125emX}}

\makeatletter
\newcommand{\linebreakand}{%
  \end{@IEEEauthorhalign}
  \hfill\mbox{}\par
  \mbox{}\hfill\begin{@IEEEauthorhalign}
}
\makeatother
\begin{document}

\title{Evaluating Self-Supervised Speech Models via Text-based LLMs
}

\name{Takashi Maekaku$^{\star}$, Keita Goto$^{\star}$, Jinchuan Tian$^{\dagger}$, Yusuke Shinohara$^{\star}$, Shinji Watanabe$^{\dagger}$}
\address{$^{\star}$ LY Corporation, Tokyo, JAPAN \\%
         $^{\dagger}$ Carnegie Mellon University, PA, USA}

\maketitle

\begin{abstract}
Self-Supervised Learning (SSL) has gained traction for its ability to learn rich representations with low labeling costs, applicable across diverse downstream tasks. 
However, assessing the downstream-task performance remains challenging due to the cost of extra training and evaluation. 
%However, evaluating SSL models' task-generalization remains challenging due to the cost of extra training and evaluation. 
Existing methods for task-agnostic evaluation also require extra training or hyperparameter tuning. We propose a novel evaluation metric using large language models (LLMs). By inputting discrete token sequences and minimal domain cues derived from SSL models into LLMs, we obtain the mean log-likelihood; these cues guide in-context learning, rendering the score more reliable without extra training or hyperparameter tuning. Experimental results show a correlation between LLM-based scores and automatic speech recognition task. Additionally, our findings reveal that LLMs not only functions as an SSL evaluation tools but also provides inference-time embeddings that are useful for speaker verification task. 
%This work introduces a new, efficient evaluation strategy without hyper-parameter tuning.
%Furthermore, integrating LLM embeddings into HuBERT improves task performance, demonstrating that LLMs can extract meaningful speach-related features. This approach offers a label-free, parameter-free evaluation method with low implementation cost, highlighting the potential of LLMs in assessing SSL models for speech processing.

%Experiments show a correlation between SSL model performance and LLM likelihood scores, and integrating LLM embeddings into HuBERT improves performance on some downstream tasks. 
%The key contribution is a label-free, no-training-needed evaluation metric leveraging LLMs, suggesting their potential as valuable features for speech downstream tasks.
\end{abstract}

%\begin{IEEEkeywords}
\vspace{-10pt}
\section{Introduction}
\vspace{-10pt}
%In recent years, self-supervised learning (SSL) has emerged as a powerful paradigm for learning high-quality, task-agnostic representations across diverse domains, including speech processing.
In recent years, self-supervised learning (SSL) has emerged as a powerful paradigm for learning high-quality, task-agnostic representations \cite{mohamed2022self}. In the speech domain, SSL encoders such as wav2vec 2.0 \cite{baevski2020wav2vec}, HuBERT \cite{hsuHubert}, WavLM \cite{chen2022wavlm}, BEST-RQ \cite{chiu2022self} and multilingual models like XLS-R \cite{babu2021xls} and XEUS \cite{chen2024towards}, and so on have pushed the state of the art on a wide range of downstream tasks.
By minimizing the reliance on labeled data, SSL has enabled the development of robust models that can be fine-tuned for various downstream tasks such as automatic speech recognition (ASR) and speech language understanding (SLU). %However, evaluating the task-generalization capability of SSL models remains a significant challenge. 
However, assessing the downstream-task performance of SSL models remains a significant challenge. 
Traditional benchmarking approaches often require extensive additional training and evaluation on various tasks such as SUPERB benchmark \cite{yang2021superb, tsai2022superb, shi23g_interspeech} and the evaluation framework based on larger-capacity probing heads \cite{zaiem2025speech}, which is both time-consuming and resource-intensive.

To address this, several methods have been proposed to estimate the performance of SSL models without additional training, such as correlation-based analysis \cite{pasad2021layer, pasad2023comparative} using canonical correlation analysis \cite{hotelling1992relations} or its variant \cite{pwcca} to compare each layer's representation to phonetic units and word meaning, and so on. 
These approaches, however, depend on precise phoneme- or word-level alignments produced by a separate model-driven forced-alignment system, and also require labeled data.
%These approaches, however, often rely on hyperparameter tuning and are limited by the availability of labeled data or model-specific information even though no additional training is required.
Reference \cite{chung2021similarity} has shown a high positive correlation between the pre-training loss of an SSL model and its downstream performance; however, that loss information is typically inaccessible unless the model has been trained in-house.
Phonetic discriminability can also be assessed with the ABX error metric \cite{schatz2013evaluating} \cite{dunbar2021zero}, but this approach demands an evaluation set specifically constructed from triplet items.
Therefore, there is a growing need for novel, label-free, parameter-free and training-free evaluation methods that can effectively capture the task-generalization potential of SSL models.
Therefore, there is a growing need for novel, label-free, parameter-free and training-free evaluation methods that can reliably gauge the cross-task transferability of SSL models.
%Therefore, there is a growing need for novel, label-free, parameter-free and training-free evaluation methods that can effectively capture the task-generalization potential of SSL models.

Inspired by the remarkable progress in text-based large language models (LLMs) \cite{brown2020language,hadi2023survey}, this paper explores the potential of leveraging these models as a new evaluation metric for speech SSL models. 
Since text-based LLMs are trained on a wide spectrum of character sequences, including natural-language text, structured data such as XML and source code, and even mathematical expressions, we hypothesize that, even without explicit training on speech, LLMs may still possess the intrinsic capability to predict a discrete token sequence that compactly encodes the information contained in speech.
%Text LLMs, trained on vast amounts of textual data, are capable of learning rich, multi-modal representations that can be applied to various tasks. Notably, they have shown the ability to predict discrete token sequences, which could be leveraged to evaluate the performance of SSL models in speech downstream tasks.

In this study, we propose a novel evaluation method that utilizes the log-likelihood of text-based LLMs when provided with discrete token sequences derived from SSL models. This approach requires no additional training or hyperparameter tuning, making it highly efficient and scalable. Our experiments demonstrate that the proposed method correlates with the performance of SSL models on ASR task. Furthermore, we demonstrate that it is possible to perform the speaker verification task using the embedding vectors obtained during LLM inference. This finding suggests that LLMs are not only useful for the ASR evaluation of SSL models, but also carry rich information that can be leveraged for the speaker verification task. This highlights the versatility and generalization capability of the proposed method.
%Furthermore, we show that incorporating the embedding representations generated by LLMs into models like HuBERT \cite{hsuHubert} can enhance performance on certain downstream tasks, highlighting the potential of text-based LLMs to extract meaningful, task-relevant information from SSL representations.

Our contributions are summarized as follows:
\begin{itemize}
    \item We propose a novel, label-free, parameter-free and training-free evaluation metric for SSL speech models, validates the utility of text-based LLMs as an ASR evaluation metric.
    \item We reveal that the embedding representation obtained during LLM inference can capture speaker-related characteristics without any additional training.
\end{itemize}
%This paper contributes a new, label-free, parameter-free and training-free evaluation metric for SSL speech models, validates the utility of text-based LLMs as an ASR evaluation metric. In addition, we reveal that the embedding representation obtained during LLM inference can capture speaker-related characteristics without any additional training.
%their potential to enhance downstream task performance in speech processing.

Although highly relevant to our study, the STAB benchmark \cite{vashishth2024stab} provides lightweight diagnostic metrics for speech tokenizers, whereas our work is the first to drive an LLM with such discretized tokens and demonstrate that their ASR performance can be ranked using the likelihood alone.

The following section reviews our proposed evaluation score followed by behavioral analysis to validate our method in Section~\ref{sec:pre_ex}. Main results are presented in Section~\ref{sec:main_ex}.

\vspace{-10pt}
\section{Proposed Method}
\vspace{-10pt}
\label{sec:proposed_method}

In this section, we describe a novel evaluation metric of SSL models using text-based LLMs designed to address the limitations of existing methods, which require additional training or hyper-parameter tuning.
Text-based LLMs, although these have not been trained on speech data, are trained on a wide spectrum of character sequences, including natural-language text, structured data such as XML and source code \cite{li2023starcoder}, and even mathematical expressions \cite{ling2017program}. We hypothesize that, even without explicit training on speech, LLMs may still possess the intrinsic capability to predict a discrete token sequence that compactly encodes the information contained in speech.

Thus, we propose to evaluate the predictability of the discrete token sequence obtained from SSL models through LLM inference by calculating the mean log-likelihood.
By comparing these likelihood scores across SSL models, we can assess how easily LLMs can predict the sequence. An SSL model that achieves a higher score can be regarded as producing discrete token sequences that contain less noise and exhibit greater grammatical and syntactic plausibility in natural language than a model with a lower score.
%First, each input speech is input to the SSL model, to obtain a discrete token sequence. 
%After deduplication, this sequence is then fed to the LLM, which returns a likelihood score.
The details of calculating our proposed metric are described below.

\subsection{Computation of the Proposed Metric}
The procedure for evaluating SSL models using LLMs given a certain speech is described below.
%\begin{itemize}
Let $\mathbf{X}$ denote an input speech waveform.  
          Feeding $\mathbf{X}$ into SSL models yields frame-level latent representations  
          \vspace{-5pt}
          \begin{equation}  
          \label{eq:x}
              \tilde{\mathbf{X}}
              =\mathrm{SSL}(\mathbf{X}),
          \end{equation}
%          \vspace{-5pt}
          where $\tilde{\mathbf{X}} = (\tilde{x}_1, \tilde{x}_2, ..., \tilde{x}_T), \tilde{x}_t \in \mathbb{R} ^D$, $T$ is the number of frames and $D$ is the feature dimension.
The latent sequence $\tilde{\mathbf{X}}$ is discretized via $k$-means clustering.  
          Each representation $\tilde{x}_t$ is assigned to its nearest centroid vector, producing a
          discrete token sequence
          \begin{equation}
          \label{eq:z}
              \mathbf{Z} = (z_1,\,z_2,\ldots,z_T), \qquad 
              z_t\in\{1,\ldots,K\},
          \end{equation}
          where $K$ is the number of clusters.
Consecutive duplicate tokens in $\mathbf{Z}$ are removed to obtain the following:
          \begin{equation}
          \label{eq:z_p}
              \tilde{\mathbf{Z}}
              = (\hat{z}_1,\,\hat{z}_2,\ldots,\hat{z}_{T'}), \qquad
              \hat{z}_{i}\neq\hat{z}_{i+1},\; T'\le T .
          \end{equation}
Then, we create an input string to LLMs. Here, we use context utterances before and after target input $\tilde{\mathbf{Z}}$ as a prompt.
%Next, we input $ \tilde{\mathbf{Z}} $ into the LLM, while providing the preceding and following utterances as context information within the prompt. 
This enables LLMs to learn properties of the discrete sequence by leveraging in-context learning with limited examples.
Example prompt and input are as follows:
%The following is an example of the prompt and the input string.
\begin{quote}
\small
\textbf{Prompt:}\\
$\verb|<prefix>|[[2\hspace{0.3em}4\hspace{0.3em}9 \cdots], [11\hspace{0.3em}4\hspace{0.3em}2 \cdots],\cdots]\verb|</prefix>|,\\ \verb|<suffix>|[[7\hspace{0.3em}2\hspace{0.3em}6 \cdots], [3\hspace{0.3em}6\hspace{0.3em}7 \cdots],\cdots]\verb|</suffix>|$ \\
\textbf{Input:}\\ $[21\hspace{0.3em}12\hspace{0.3em}1\hspace{0.3em}9\hspace{0.3em}83\hspace{0.3em}\cdots]$
\end{quote}
%In the above example, the prompt includes both the context sequence that comes immediately before the target sequence and the one that comes immediately after it. 
Tags such as ``\texttt{<prefix>}" and ``\texttt{<suffix>}" is a syntactic marker that indicates whether the accompanying sequence precedes or follows the target. %In the prompt we insert guide tokens such as ``\texttt{<prefix>}" to explicitly mark whether the accompanying context precedes or follows the target sequence.
Let $\mathbf{S}_n$ denote the $n$-th input sequence to LLMs, we define the set of all sequences as $\textbf{S}=\{\textbf{S}_1, \textbf{S}_2,\ldots,\textbf{S}_N\}$, where $N$ is the number of utterances.
%Let $\mathbf{S}$ denote the full input sequence to the LLM, 
$\textbf{S}_n$ is composed of the prompt $\mathbf{S}_n^{\texttt{(pr)}}$ and the target input $\mathbf{S}_n^{\texttt{(in)}}$, i.e.,\ $\mathbf{S}_n=(\mathbf{S}_n^{\texttt{(pr)}},\mathbf{S}_n^{\texttt{(in)}})$.
Note that in this study, $\mathbf{S}_n$ is treated as just a string. This eliminates the need to add a new LLM token corresponding to the discrete token, and no additional LLM training is required.

The mean log-likelihood (MLL) over the input part of the sequence can be calculated as follows:
\begin{equation}
    \text{MLL}(\mathbf{S};\theta) =
    \frac{1}{\sum_{n}T'_n}\sum_{t,n}
    \log p_{\theta}\!\bigl(s^{\texttt{(in)}}_{t+1,n}\mid s^{\texttt{(in)}}_{\le t,n},\mathbf{S}_n^{\texttt{(pr)}}\bigr).
    \qquad
\end{equation}

%\begin{equation}
%    \text{MLL}(\mathbf{S};\theta) \;=\;
%    \frac{1}{NL}\sum_{i=1}^{N}\sum_{t=1}^{L}
%    \log p_{\theta}\!\bigl(s^{\texttt{(in)}}_{t+1,i}\mid %s^{\texttt{(in)}}_{\le t,i}\bigr).
%    \qquad
%\end{equation}
%\begin{equation}
%    \text{MLL}(\mathbf{x};\theta) \;=\;
%    \frac{1}{T-P}\sum_{t=P}^{T-1}
%    \log p_{\theta}\!\bigl(x_{t+1}\mid x_{\le t}\bigr).
%    \qquad
%\end{equation}
where $\mathbf{S}_n^{\texttt{(in)}} = (s^\texttt{(in)}_{1,n},\,s^\texttt{(in)}_{2,n},\ldots,s^\texttt{(in)}_{T'_{n},n})$ and $\theta$ represents the parameters of LLMs.
In particular, we refer to the score before averaging over the entire dataset as the \textit{per-utterance MLL}, which will be used in Section~\ref{sec:main_ex}.

%\end{itemize}

The actual evaluation is done by calculating the MLL of our target speech database for each of the different SSL models and comparing the relative difference in value. Since this MLL score is a metric focusing on the way discrete tokens transition, it is inferred to be highly relevant to the ASR task, but specific verification is provided in Section~\ref{sec:main_ex}.
%In the next section, we verify whether this evaluation metric can in fact be applied to speech data through behavioral analysis.
\vspace{-10pt}
\section{behavioral analysis}
\label{sec:pre_ex}

This section investigates whether the proposed evaluation metric behaves as intended in a series of pilot studies. %Three preliminary experiments are conducted. 
We first examine how the MLL varies as the length of the input sequence increases. We then measure the effect of supplying additional context—tokens that precede and follow the target sequence—as part of the prompt. Although LLMs have never been trained on speech data, a consistent increase in the MLL score with longer sequences or richer context would suggest that the LLM captures the grammatical or syntactic regularities present in the input. Finally, we explore the MLL score’s sensitivity to different prompt patterns.

\subsection{Experimental Setup}

We employed Gemma3-4B \cite{kamath2025gemma}\footnote{Although precisely this 4B model is a multi-modal LLM with a vision encoder, all analyses in this chapter confirmed the same trend for the 1B model, which is trained only on text. Evaluation using other text-based LLMs will be described in the next chapter.} as the LLM for calculating the MLL.
The test speech data consisted of 100 utterances randomly selected from the 100hrs subset of LibriSpeech \cite{panayotov2015librispeech} (“train-clean-100’’).
Timestep-level speech representations as in Eq.~(\ref{eq:x}) were extracted at 50 frames per second using HuBERT-Base that had been trained for two iterations on the same 100 hrs subset.
To convert HuBERT features into discrete tokens as in Eq.~(\ref{eq:z}), we trained a $k$-means on 10\% of the train-clean-100 and used the resulting centroids to quantize every time step. The number of clusters $K$ is 500.
The resulting token sequences were deduplicated as in Eq.~(\ref{eq:z_p}) then fed into Gemma3 for the inference, which was carried out with Hugging Face \cite{jain2022hugging}.

\subsection{Variation of MLL when input string length is varied}

First, we investigated how the MLL scores varied when the length of the input string was varied. The results are shown in Table~\ref{tab:series_len}, where ``Target Token Length" is the number of characters cut out from the beginning of the input, not the number of tokens.
``Max" is the case where the entire input string was entered, in which case the average string length was around 1400 characters.
The results show that the longer the input sequence becomes, the easier the prediction becomes.
This suggests that longer input sequences supply richer context information from the past, making next-token prediction progressively easier.
Based on this, we can infer that the speech data has a high degree of statistic consistency with the LLM.
%As can be seen in Table~\ref{tab:series_len},

\begin{table}[t]
  \centering
  \footnotesize
  \caption{MLL obtained when varying the length of the input string.
           A length of 500 indicates that only the first 500 characters of the input token sequence is used, whereas \textit{Max} indicates that the entire
           sequence is processed.}
  \label{tab:series_len}
  \adjustbox{max width=0.9\textwidth}{
  \begin{tabular}{lc}
    \toprule
    Target Character Length & MLL \\ \midrule
    500  & -1.610 \\ 
    1000 & -1.588 \\ 
    Max  & -1.568 \\ \bottomrule
  \end{tabular}
  }
\end{table}

\begin{table*}[t]
  \centering
  \footnotesize
  \caption{Evaluation results for each prompt pattern. $X^-$ and $X^+$ denote the contextual utterances in the past and future directions, respectively.}
  \label{tab:prompt_patterns}
   \adjustbox{max width=0.9\textwidth}{
  \begin{tabular}{clc}
    \toprule
    No.\ & Prompt Pattern                               & MLL \\ \midrule
     1   & \texttt{Past:} [$X^-$], \texttt{Future:} [$X^+$]            & -1.580  \\
     2   & \texttt{Past Utterances:} [$X^-$], \texttt{Future Utterances:} [$X^+$]               & -1.588  \\
     3   & \verb|<prefix>|[$X^-$]\verb|</prefix>|, \verb|<suffix>|[$X^+$]\verb|</suffix>|        & \textbf{-1.568}  \\
     4   & \verb|<past_utterances>|[$X^-$]\verb|</past_utterances>|, \verb|<future_utterances>|[$X^+$]\verb|</future_utterances>|                         & -1.578  \\
     5   & \verb|<previous_utterances>|[$X^-$]\verb|</previous_utterances>|, \verb|<subsequent_utterances>|[$X^+$]\verb|</subsequent_utterances>|          & -1.576  \\
     6   & \verb|<past_context>|[$X^-$]\verb|</past_context>|, \verb|<future_context>|[$X^+$]\verb|</future_context>|       & -1.570  \\
     7   &  \verb|<prefix_sequences>|[$X^-$]\verb|</prefix_sequences>|, \verb|<suffix_sequences>|[$X^+$]\verb|</suffix_sequences>|    & -1.574  \\ \bottomrule
  \end{tabular}
  }
\end{table*}

\begin{table}[!t]
    \caption{Pearson correlation coefficient between the per-utterance MLLs obtained with discrete tokens from final layer of HuBERT Base and those obtained with the transcription.}
    \centering

    \label{tab:spearman}
    \begin{adjustbox}{width=0.7\linewidth}
    % ---------------- no vertical rules, three horizontal rules -------------
    \begin{tabular}{lcc}
        \toprule
        \textbf{Method} & \textbf{dev-clean} & \textbf{dev-other} \\

        \midrule
        $\rho$\,(Gemma3-4B, Trans.)                       & 0.547 & 0.487 \\
        $\rho$\,(Qwen3-4B, Trans.)                       & 0.437 & 0.363 \\
        $\rho$\,(Phi-4-mini, Trans.)                       & 0.497 & 0.410 \\
        \bottomrule
    \end{tabular}
    \end{adjustbox}
\end{table}

%\subsection{Effect of Context Size on the Proposed Score}
\subsection{Impact of Varying Information Density in In-Context Learning}
In this analysis, we investigated how the MLL changes when the context size of $\mathbf{S^{\texttt{(pr)}}}$
%the amount of contextual information surrounding the target sequence 
is gradually increased. Since LibriSpeech corpus is read speech and provides chapter annotations, we selected the utterances that immediately precede and follow the target utterance within the same chapter, expanding the context size symmetrically (e.g., $\pm 1$, $\pm 2$, $\pm 3$ utterances). When a target utterance occurred near the beginning or the end of a chapter and the required number of neighboring utterances was unavailable, the missing context slots were filled with the string ``\texttt{N/A}" in the prompt so that the overall prompt format remained constant across all conditions.

\begin{figure}[htbp]
\centerline{\includegraphics[clip,width=5.0cm]{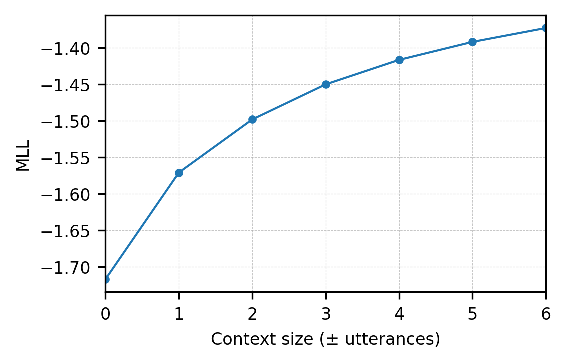}}
\caption{Relationship between context size (number of preceding and succeeding utterances) and the MLL.}
\label{fig:context}
\end{figure}

The results are shown in Figure~\ref{fig:context}. \footnote{Note that due to the memory limit of the GPU environment used in the experiment, the context size was limited to 6.}As we can see, the MLL score increases monotonically as the context size increases. From this, we can interpret that the model is effectively utilizing the information of the added context and that the understanding of the speech domain data is increasing. Therefore, this indicates that the addition of the context allows the LLM to evaluate the naturalness of the sequences more suited to the speech domain of the input data through in-context learning without any additional training.

\vspace{-5pt}
\subsection{Influence of Prompt Template Variations}

To investigate the sensitivity of the proposed metric to the wording and structure of the prompt itself, we systematically varied the prompt template and measured the resulting MLL differences.  As shown in Table~\ref{tab:prompt_patterns}, seven distinct templates were prepared. $X^-$ and $X^+$ denote the context sequences in the past and future directions, respectively. The templates differ along several linguistic and formatting dimensions: (i) how the notions of ``past" and ``future" context are paraphrased, (ii) whether or not XML-style tags are used to delimit the contextual segments, and (iii) whether explicit cues such as ``utterance" or ``sequence" are included to describe the data type.  
%\enlargethispage{-10pt}
From the results, comparing No.~2 and No.~4, it is clear that the tagged patterns have higher scores. Also, a comparison of Nos.~4--7 shows that MLL did not increase even if the template explicitly states that the input is an utterance. This is a natural result since the training data does not include speech.
The pattern in No.~3 had the highest score. From this, it can be said that expressions that the model is familiar with at the time of training are more effective in helping the model understand than what kind of sequence the input data is.
\enlargethispage{-20pt}  
However, the variation in MLLs is small compared to the results for different context sizes of $\mathbf{S^{\texttt{(pr)}}}$. Therefore, it can be said to be robust to such fluctuations in the template pattern.
In subsequent experiments, the No.~3 template will be used unless otherwise noted.

From the above behavioral analyses, it can be said that the LLM attempts to capture the statistical characteristics of the input data to some extent, although the LLM does not recognize the input token sequences as data derived from the speech domain.
%it is capable of capturing the contextual and statistical properties of this data.
These analysis consistently indicates that the MLL score improves monotonically with both the length of the input sequence and the size of the prompt context; in other words, longer sequences and richer contextual information are always beneficial within the range we tested. In contrast, the choice of prompt template pattern exerts only a marginal influence and does not materially alter the outcome. Consequently, no model-specific tuning appears to be required: practitioners can simply select the context size in accordance with their computational budget, secure in the knowledge that the procedure does not hinge on any sensitive hyper-parameter.
%Note that multiple prompts and context sizes were explored solely to verify that feeding discrete token sequences from speech domain into LLM produces valid scores; \textit{they should therefore not be regarded as hyper-parameters}. 
Section~\ref{sec:main_ex} further demonstrates that the 
proposed metric remains effective even when the target sequence is evaluated without its immediate preceding and succeeding context sequences, provided that a single example sequence is supplied alongside it.
%proposed metric remains effective even when the sequence is evaluated without its immediate preceding and succeeding context sequences.
%\vspace*{-10pt}
\section{Main Experiments}
%\vspace*{-10pt}
\label{sec:main_ex}
Building on the findings of the previous section—which confirmed the validity of estimating the MLL score of token sequence from SSL models by means of LLMs—we now examine how this score correlates with SSL model’s effectiveness on a downstream task. Since the MLL score is designed to capture the statistical naturalness of token transitions, ASR is chosen as the evaluation task.
%ASR task should benefit directly from more linguistically plausible sequences. 
In addition, we conduct a comparative analysis employing several alternative LLMs and other SSL models, to determine how the choice of these models influences the observed relationship.
%\balance 
%\vspace*{-20pt}
\begin{figure*}[htbp]
\centerline{\includegraphics[clip,width=11.0cm]{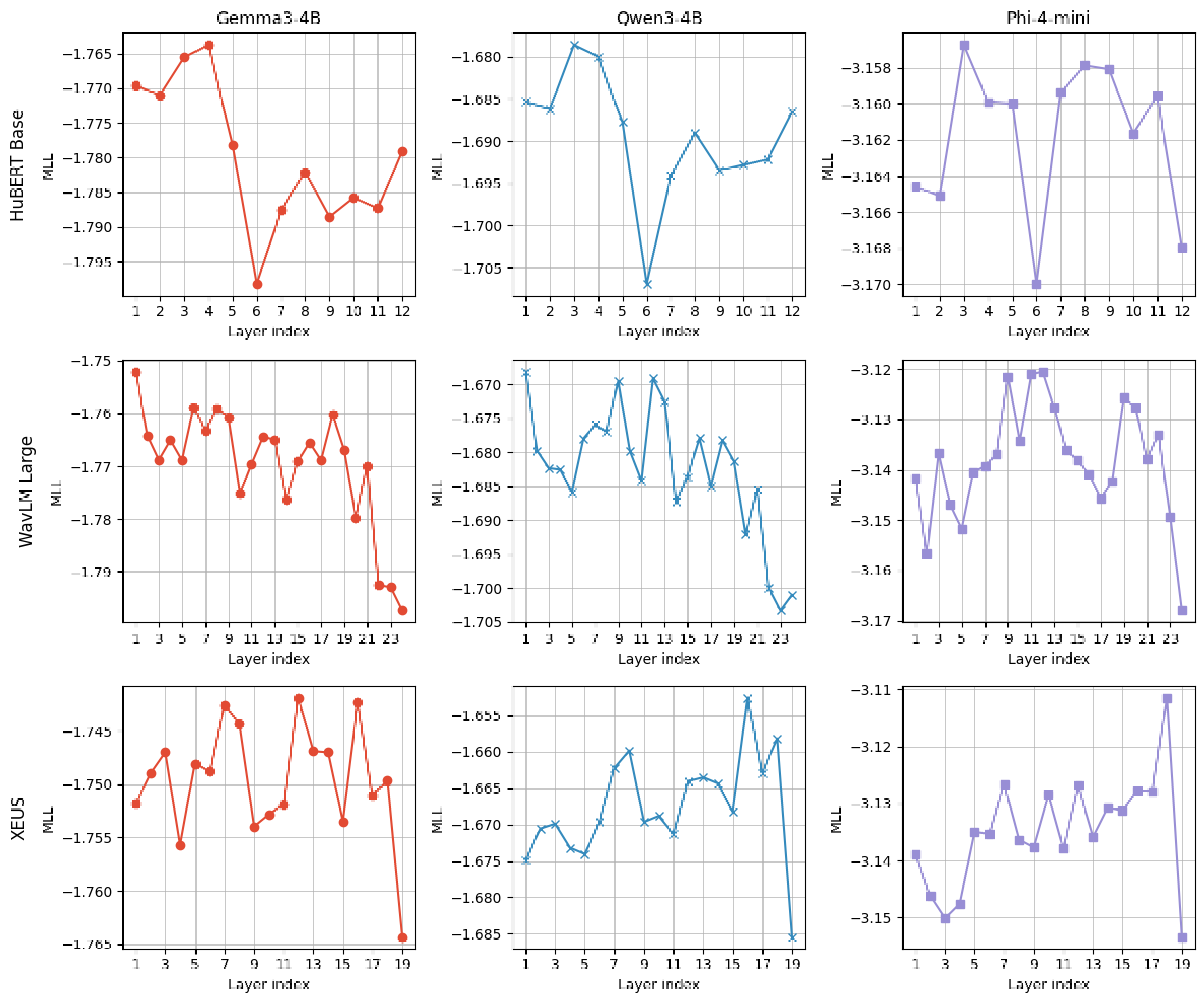}}
\caption{Layer-wise MLL comparison for the three SSL models.
MLLs are computed from the outputs of three LLMs (Gemma3-4b, Qwen3-4b, Phi-4-mini).
The left y-axis reports scores obtained with Phi-4-mini, whereas the right y-axis reports scores obtained with sub-models Gemma3-4b and Qwen3-4b.}
\label{fig:mll}
\end{figure*}
\vspace*{-7pt}
\subsection{Experimental Setup}

\subsubsection{Models}
\vspace*{-7pt}
In addition to Gemma3-4B, Qwen3-4B \cite{yang2025qwen3} and Phi-4-mini (3.8B) \cite{abouelenin2025phi} were employed to compute the MLL score.  Both are trained exclusively on textual data that include natural-language documents, large-scale web corpora, and source code. Qwen3-4B is notable for its coverage of 119 languages and dialects, whereas Phi-4-mini distinguishes itself by incorporating high-quality synthetic data such as mathematical problems and source code into its training corpus.

Besides HuBERT Base that had been trained for two iterations
on the LibriSpeech 100 hrs subset, two additional HuBERT-style SSL models were examined: WavLM Large and XEUS.
WavLM performs masked prediction task while simultaneously denoising speech that has been artificially corrupted with environmental noise.
XEUS extends this strategy by incorporating dereverberation in addition to denoising and is trained on an extremely multilingual dataset that covers 4,057 languages.
%The number of clusters $K$ is 500 for all SSL models.
$K$ is 500 for all SSL models.

\subsubsection{ASR training}
\vspace*{-7pt}
Finetuning and evaluation of the ASR task were conducted within the SUPERB \cite{yang2021superb} framework.
All hidden layers of each SSL model encoder are combined through a trainable weighted sum rather than relying on the last layer alone, and this composite representation is passed to a bidirectional LSTM \cite{graves2005framewise} with two 1024-unit layers trained with CTC \cite{graves2012connectionist} on character targets. Decoding is carried out by beam search using the official LibriSpeech four-gram language model implemented with KenLM \cite{heafield2011kenlm} and the Flashlight \cite{pratap2019wav2letter++} toolkit. 
%All hyper-parameter details are omitted here for brevity.
%During HuBERT’s ASR fine-tuning stage, we optimize the model with the connectionist temporal classification (CTC) \cite{graves2012connectionist} objective. The original projection head is removed and replaced by a newly initialized soft-max layer, and all experiments are implemented in the ESPnet toolkit \cite{watanabe2018espnet}.
%Only ``train-clean-100" from LibriSpeech was used for the fine tuning data.
%For WavLM's fine-tuning, they attach a randomly initialized linear head to the Transformer encoder, train with a CTC objective using character-level targets, and apply SpecAugment \cite{park2019specaugment} style time and channel masking together.
%and a tri-stage Adam learning-rate schedule on multi-GPU batches. 
%Further architectural choices and hyper-parameter settings can be found in \cite{chen2022wavlm}.
%As for XEUS, the downstream probe employs a lightweight two-layer Transformer encoder trained with the CTC objective. 
%Each task is run for a predetermined number of steps using the Adam optimizer (Kingma and Ba, 2015) with a constant learning-rate schedule, where the learning rate is the sole hyperparameter we sweep over a small range. 
%For full architectural and hyper-parameter details, please refer to \cite{chen2024towards}.
%\subsubsection{MLL Setting}

%In these experiments except for the last one, the context size of the sequences described in the prompt was one utterance before and after.

\vspace*{-7pt}
\subsection{Results}
\vspace*{-3pt}
\subsubsection{Correlation between per-utterance MLLs obtained from discrete tokens and those obtained from transcriptions}
%Before reviewing the connection between ASR performance and MLL, 
We first investigated the relationship between per-utterance MLL scores obtained from discrete token-sequence inputs from final layer of HuBERT and those obtained from transcription inputs in Table~\ref{tab:spearman}\footnote{Incidentally, when MLL (averaged over the entire data) is calculated using the discrete token sequences as input and the transcribed text as input, MLL is considerably larger in the former case. For example, when Gemma3-4B was used for LLM and ``dev-clean" for the data set, the former was -1.615 and the latter was -4.421. This is because LLM is not fine-tuned with any newly added tokens, but simply inputs the token sequence as a string of characters, so the prediction is for a total of 11 different characters (numbers from 0 to 9 and spaces), and the relative vocabulary is much smaller.}. 
%In the upper block, MLLs are given as ``A / B", where the left‐hand number is the score computed with discrete token sequences and the right‐hand number is the score computed with the transcription. The lower block lists the Pearson correlation coefficient between the per-utterance MLL obtained with discrete token sequences and those obtained with the transcription.
As can be seen, all models exhibit a moderate positive correlation between the two conditions. Although the correlation is lower for the ``dev-other", this can be attributed to the degraded speech quality cased by noise and other artifacts, which likely results in noisier token sequences. Therefore, LLM appears to produce the per-utterance MLL with similar relative magnitudes for both discrete tokens and the transcriptions. 
This implies it captured common grammatical, lexical-frequency, or character-transition patterns between the inputs.
%It suggests that it may have captured shared grammatical structures, frequent word distributions, or character transition patterns across the two inputs.

\subsubsection{Relationship Between ASR performance and MLL}
We analysed the connection between the MLL and word error rate (WER) obtained with three SSL models.  
For every layer of each encoder, the latent features were first discretised; MLL was then computed on the resulting token sequence and finally averaged across all layers. 
We varied the LLM among the three models introduced in Section~\ref{sec:main_ex}-A.
%The LLM was varied among the three LLMs introduced earlier, i.e., Gemma3-4B, Qwen3-4B, and Phi-4-mini to check the robustness of the observation.

In addition to the setup that uses the single discrete token sequence immediately preceding the target sequence and the single sequence immediately following it as context, an alternative condition is evaluated in which the initial sequence in the same chapter is selected once and reused as the prompt for every sequence in that chapter. This setting tests whether supplying a data sample—without an explicit temporal context—enables LLMs to capture the input characteristics in an in-context manner and to apply the MLL as effectively as in the explicit-context condition.

\begin{figure*}[htbp]
\centerline{\includegraphics[clip,width=14.5cm]{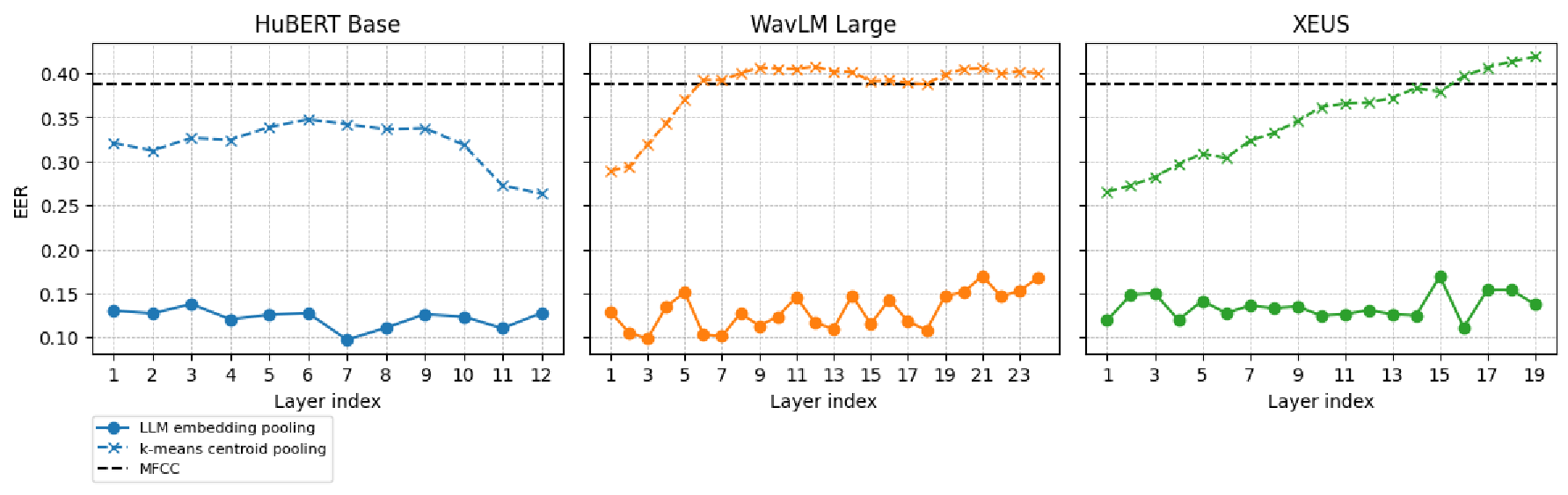}}
\caption{EER comparison across layers of three SSL models. The final-layer hidden representations from Gemma3-4B are used to perform SV task, and the resulting EER is recorded.}
\label{fig:sv}
\end{figure*}

ASR performance was measured on the ``test-clean" set of LibriSpeech, while the MLL was computed on a random 10 \% subset of the ``test-clean" in order to reduce evaluation cost. Table~\ref{tab:model_comparison} summarizes the results. ``P1" refers to the case where the discrete token sequences immediately before and after the target sequence are used as the prompt ($\pm 1$ utterances as the context), while ``P2" uses a sequence that corresponds to the first utterance of the same chapter as the prompt as mentioned in the previous paragraph. Regardless of which LLM was used or which prompt pattern was applied, the encoder that achieved the lowest WER, i.e., XEUS also yielded the highest MLLs, whereas the encoder with the highest WER, i.e., HuBERT produced the lowest MLLs. This concordance indicates that the MLL is strongly correlated with ASR performance and can serve as a proxy for comparing SSL models without any task-specific fine-tuning.
In addition, the MLL found to be robust to variations in prompt patterns. In particular, the results obtained using the prompt from ``P2" suggests that the MLL remains effective even in scenarios where contextual information is not necessarily available.
%\footnote{The discrete token sequence fed to the LLM is obtained by k-means clustering, whose cluster indices are arbitrary; any one-to-one relabeling (e.g., the sequence “1 2 1 5 5” permuted to “21 52 21 8 8”) should, in principle, encode the same information.
%Although a rigorous study would evaluate performance under such index permutations, we observe essentially the same trends across three independent LLM backbones, suggesting that the influence of arbitrary label permutations is negligible for the experiments.}

Figure~\ref{fig:mll} plots layer-wise MLL curves for every combination of SSL models and LLMs. In contrast to existing methods using CCA \cite{pasad2021layer,pasad2023comparative} that rely on labeled alignments, our approach provides label-free insights into the internal representation of SSL models. As we can see, although previous studies report that higher layers of WavLM contribute most to ASR performance \cite{chen2022wavlm}, MLL does not always peak in the same layers. This indicates that LLMs consider the discrete token sequence to be statistically natural to some degree, whether the sequence is obtained from the lower or higher layers. When comparing LLMs within the same SSL model, Gemma3-4B and Qwen3-4B exhibit almost identical trajectories
%(differing in absolute scale due to their entry size of the tokenizer
%\footnote{Tokenizer vocabulary sizes: Gemma3-4B = 262k, Phi-4-mini = 200k, Qwen3-4B = 150k entries.} )
, whereas Phi-4-mini exhibits a little different pattern. This difference may stem from Phi-4-mini’s training corpus, which emphasizes synthetic data such as mathematical problems and source-code \cite{abouelenin2025phi}, thereby altering its MLL statistics.

\begin{table}[tb]
  \caption{ASR scores for three models on a 10\% subset of \textit{test-clean}. ``P1" refers to the case where the discrete token sequences immediately before and after the target sequence are used as the prompt, while ``P2" uses a sequence that corresponds to the first utterance of the same chapter as the prompt.}
  \label{tab:model_comparison}
  \centering
  \resizebox{0.8\linewidth}{!}{
  \setlength{\tabcolsep}{3pt}
  \begin{tabular}{lccccccc}
    \toprule
    \multirow{3}{*}{\textbf{Model}} &
    \multirow{3}{*}{\textbf{WER}\,$\downarrow$} &
    \multicolumn{6}{c}{\textbf{MLL}\,$\uparrow$} \\ \cmidrule(lr){3-8}
    & & \multicolumn{2}{c}{\textbf{Gemma3-4B}} &
        \multicolumn{2}{c}{\textbf{Qwen3-4B}} &
        \multicolumn{2}{c}{\textbf{Phi-4-mini}} \\ \cmidrule(lr){3-4}\cmidrule(lr){5-6}\cmidrule(lr){7-8}
    & & \textbf{P1} & \textbf{P2} &
        \textbf{P1} & \textbf{P2} &
        \textbf{P1} & \textbf{P2} \\
    \midrule
    HuBERT & 14.78 & -1.780 & -1.731 & -1.689 & -1.668 & -3.162 & -3.190 \\
    WavLM & 3.44 & -1.770 & -1.712 & -1.683 & -1.647 & -3.138 & -3.139 \\
    XEUS & \textbf{3.34} & \textbf{-1.750} & \textbf{-1.677} &
                     \textbf{-1.668} & \textbf{-1.613} &
                     \textbf{-3.135} & \textbf{-3.101} \\
    \bottomrule
  \end{tabular}
  }
\end{table}
\begin{comment}

\begin{table}[tb]
  \caption{ASR Performance comparison of three models. Scores are computed on a 10\% subset of \textit{test-clean}, with one preceding utterance and one succeeding utterance provided as context in the prompt.}
  \label{tab:model_comparison}
  \centering
  % The IEEE style discourages vertical rules; booktabs gives clean horizontals.
  \begin{tabular}{lccccccc}
    \toprule
    \multirow{3}{*}{\textbf{Model}} &
    \multirow{3}{*}{\textbf{WER}} &
    \multicolumn{6}{c}{\textbf{MLL}} \\
    \cmidrule(lr){3-8}
    & & \textbf{Gemma3-4b} & \textbf{Qwen3-4b} & \textbf{Phi-4-mini} \\
        \cmidrule(lr){3-4}\cmidrule(lr){5-6}\cmidrule(lr){7-8}
    & & \textbf{P1} & \textbf{P2} & \textbf{P1} & \textbf{P2} & \textbf{P1} & \textbf{P2} \\
    \midrule
    HuBERT Base & 14.78 & -1.780 / -1.731 & -1.689 / -1.66764 & -3.162 / -3.18957 \\
    WavLM Large & 3.44 & -1.770  & -1.683 & -3.138 \\
    XEUS & 3.34 & \textbf{-1.750} & \textbf{-1.668} & \textbf{-3.135} \\
    \bottomrule
  \end{tabular}
\end{table}
\end{comment}

\subsubsection{Assessing Speaker Verification Capability}

To further explore the utility of LLMs, we conduct an auxiliary experiment on the speaker verification task by using the embedding from LLMs through inference.

%Dataset
We evaluated our method on the speaker-verification task using the VoxCeleb1 \cite{nagrani2020voxceleb} test partition. Although the original test set contains 4,874 utterances from 40 speakers, we randomly sampled a balanced subset of 20 speakers (10 male, 10 female). For each selected speaker, 10 utterances were chosen at random, yielding 200 utterances in total.
Within the 10 utterances of every speaker, one utterance was reserved as a prompt for computing the MLL score. The remaining 9 utterances served as verification trials.
For every utterance we extracted a 2,560-dimensional embedding from the last hidden layer through the LLM inference. Each embedding was compressed to 128 dimensions using principal component analysis (PCA) and subsequently $\ell_2$-normalized.
Among the nine verification utterances, one was further selected per speaker as the enrollment utterance; its frame-level embeddings were averaged over time to obtain the speaker embedding.
The evaluation set therefore consisted of $20\texttt{speakers}\times9=180$ verification utterances. Performance was reported in terms of the Equal Error Rate (EER), computed on all target and impostor trials generated from the 180 verification utterances.
%Baseline systems
Baseline systems are as follows:
(i) \emph{MFCC}, where 80-dimensional frame-level MFCCs are averaged over each utterance to obtain an 80-D speaker embedding; and (ii) \emph{Centroid-pool}, in which every discrete token is replaced by its $k$-means centroid, the centroid sequence is temporal-mean-pooled, and the resulting vector is projected to 128 dimensions via PCA.
%\textbf{Baseline systems.}
\begin{comment}
\begin{itemize}[leftmargin=*]
  \item \textbf{MFCC.} An 80-dimensional Mel-frequency cepstral coefficient (MFCC) vector, obtained by mean-pooling frame-level MFCCs over the whole utterance, served as the speaker representation.
  \item \textbf{Centroid-pool.} Each discrete token is replaced by the centroid of its $k$-means cluster; the resulting centroid sequence is mean-pooled over time and finally reduced to 128 dimensions with PCA to form the utterance-level embedding.

\end{itemize}

\begin{itemize}
    \item Baseline 1 (MFCC): An 80-dimensional Mel-frequency cepstral coefficient (MFCC) vector, obtained by mean-pooling frame-level MFCCs over the whole utterance, served as the speaker representation.
    \item Baseline 2 (Centroid pool): For each token in the discrete token sequence, it was replaced by the centroid of its assigned cluster of $k$-means, and the utterance-level representation was formed by mean-pooling these centroid vectors, and subsequently reduced to 128 dimensions using PCA.
\end{itemize}
\end{comment}

Figure~\ref{fig:sv} presents the evaluation results across three different SSL models.
Using the LLM embedding achieved a markedly lower EER than both baseline systems, indicating that the LLM embeddings captured speaker-specific characteristics more effectively.
Moreover, a layer-wise analysis revealed that LLM embeddings exhibit a smaller variance in performance across layers than the baseline methods, indicating greater robustness of the representation.
These findings provide new insights into the nature of the LLM embeddings.

\section{Conclusion}
\label{sec:conclusion}
This paper proposed a novel evaluation metric of SSL models using LLMs. 
%By inputting discrete token sequences derived from speech SSL models into LLMs, we compute the average log-likelihood, which reflects the model's ability to predict audio tokens without additional training or hyperparameter tuning. 
By simply feeding LLMs with discrete token sequences together with minimal domain cues, we can compute the mean log-likelihood in a fully label-free manner; these cues steer in-context learning and make the score more reliable, all without any additional training or hyper-parameter tuning.
Experimental results have shown that a correlation between LLM-based scores and an automatic speech recognition task.
Furthermore, our additional experiment revealed that LLMs can provide valuable
features for speaker verification task.
%Future work includes (i) extending the evaluation to additional corpora to confirm the robustness of the current findings and (ii) having confirmed that the embeddings enable the speaker verification task to be solved to a certain extent, conducting analogous tests on other downstream tasks.

%\newpage
\clearpage
\section{References}
{
\printbibliography

@ARTICLE{hsuHubert,

  author={Hsu, Wei-Ning and Bolte, Benjamin and Tsai, Yao-Hung Hubert and Lakhotia, Kushal and Salakhutdinov, Ruslan and Mohamed, Abdelrahman},

  journal={IEEE/ACM TASLP}, 

  title={{HuBERT}: Self-Supervised Speech Representation Learning by Masked Prediction of Hidden Units}, 

  year={2021},

  }

@article{nagrani2020voxceleb,
  title={Voxceleb: a large-scale speaker identification dataset},
  author={Nagrani, Arsha and Chung, Joon Son and Zisserman, Andrew},
  journal={arXiv preprint arXiv:1706.08612},
  year={2017}
}

@incollection{jain2022hugging,
  title={Hugging face},
  author={Jain, Shashank Mohan},
  booktitle={Introduction to transformers for NLP: With the hugging face library and models to solve problems},
  pages={51--67},
  year={2022},
  publisher={Springer}
}

@article{kamath2025gemma,
  title={{Gemma 3 Technical Report}},
  author={Kamath, Aishwarya and Ferret, Johan and Pathak, Shreya and Vieillard, Nino and Merhej, Ramona and Perrin, Sarah and Matejovicova, Tatiana and Ram{\'e}, Alexandre and Rivi{\`e}re, Morgane and Rouillard, Louis and others},
  journal={arXiv preprint arXiv:2503.19786},
  year={2025}
}

@inproceedings{panayotov2015librispeech,
  title={{LibriSpeech}: an {ASR} corpus based on public domain audio books},
  author={Panayotov, Vassil and Chen, Guoguo and Povey, Daniel and Khudanpur, Sanjeev},
  booktitle={Proc. ICASSP},
  pages={5206--5210},
  year={2015}
}

@inproceedings{tsai2022superb,
  title={{SUPERB-SG}: Enhanced Speech processing Universal PERformance Benchmark for Semantic and Generative Capabilities},
  author={Tsai, Hsiang-Sheng and Chang, Heng-Jui and Huang, Wen-Chin and Huang, Zili and Lakhotia, Kushal and Yang, Shu-wen and Dong, Shuyan and Liu, Andy and Lai, Cheng-I and Shi, Jiatong and others},
  booktitle={Proc. ACL},
  pages={8479--8492},
  year={2022}
}

@inproceedings{yang2021superb,
  author={Shu-Wen Yang and Po-Han Chi and Yung-Sung Chuang and Cheng-I Jeff Lai and Kushal Lakhotia and Yist Y. Lin and Andy T. Liu and Jiatong Shi and Xuankai Chang and Guan-Ting Lin and Tzu-Hsien Huang and Wei-Cheng Tseng and Ko-tik Lee and Da-Rong Liu and Zili Huang and Shuyan Dong and Shang-Wen Li and Shinji Watanabe and Abdelrahman Mohamed and Hung-yi Lee},
  title={{{SUPERB}: Speech Processing Universal PERformance Benchmark}},
  year=2021,
  booktitle={Proc. INTERSPEECH},
  pages={1194--1198},
  %doi={10.21437/Interspeech.2021-1775}
}

@article{yang2025qwen3,
  title={Qwen3 technical report},
  author={Yang, An and Li, Anfeng and Yang, Baosong and Zhang, Beichen and Hui, Binyuan and Zheng, Bo and Yu, Bowen and Gao, Chang and Huang, Chengen and Lv, Chenxu and others},
  journal={arXiv preprint arXiv:2505.09388},
  year={2025}
}

@article{abouelenin2025phi,
  title={Phi-4-mini technical report: Compact yet powerful multimodal language models via mixture-of-loras},
  author={Abouelenin, Abdelrahman and Ashfaq, Atabak and Atkinson, Adam and Awadalla, Hany and Bach, Nguyen and Bao, Jianmin and Benhaim, Alon and Cai, Martin and Chaudhary, Vishrav and Chen, Congcong and others},
  journal={arXiv preprint arXiv:2503.01743},
  year={2025}
}

@inproceedings{shi23g_interspeech,
  author={Jiatong Shi and Dan Berrebbi and William Chen and En-Pei Hu and Wei-Ping Huang and Ho-Lam Chung and Xuankai Chang and Shang-Wen Li and Abdelrahman Mohamed and Hung-yi Lee and Shinji Watanabe},
  title={{ML-SUPERB: Multilingual Speech Universal PERformance Benchmark}},
  year=2023,
  booktitle={Proc. INTERSPEECH},
  pages={884--888},
  %doi={10.21437/Interspeech.2023-1316}
}

@article{graves2012connectionist,
  title={Connectionist temporal classification},
  author={Graves, Alex},
  journal={Supervised sequence labelling with recurrent neural networks},
  pages={61--93},
  year={2012},
  publisher={Springer}
}

@article{chen2022wavlm,
  title={{WavLM}: Large-scale self-supervised pre-training for full stack speech processing},
  author={Chen, Sanyuan and Wang, Chengyi and Chen, Zhengyang and Wu, Yu and Liu, Shujie and Chen, Zhuo and Li, Jinyu and Kanda, Naoyuki and Yoshioka, Takuya and Xiao, Xiong and others},
  journal={JSTSP},
  year={2022},
  publisher={IEEE}
}

@article{chen2024towards,
  title={Towards robust speech representation learning for thousands of languages},
  author={Chen, William and Zhang, Wangyou and Peng, Yifan and Li, Xinjian and Tian, Jinchuan and Shi, Jiatong and Chang, Xuankai and Maiti, Soumi and Livescu, Karen and Watanabe, Shinji},
  journal={arXiv preprint arXiv:2407.00837},
  year={2024}
}

@inproceedings{baevski2020wav2vec,
  title={{wav2vec} 2.0: A Framework for Self-Supervised Learning of Speech Representations},
  author={Baevski, Alexei and Zhou, Yuhao and Mohamed, Abdelrahman and Auli, Michael},
  booktitle={Proc. NeurIPS},
  year={2020}
}

@article{babu2021xls,
  title={XLS-R: Self-supervised cross-lingual speech representation learning at scale},
  author={Babu, Arun and Wang, Changhan and Tjandra, Andros and Lakhotia, Kushal and Xu, Qiantong and Goyal, Naman and Singh, Kritika and Von Platen, Patrick and Saraf, Yatharth and Pino, Juan and others},
  journal={arXiv preprint arXiv:2111.09296},
  year={2021}
}

@article{graves2005framewise,
  title={Framewise phoneme classification with bidirectional LSTM and other neural network architectures},
  author={Graves, Alex and Schmidhuber, J{\"u}rgen},
  journal={Neural networks},
  volume={18},
  number={5-6},
  pages={602--610},
  year={2005},
  publisher={Elsevier}
}

@inproceedings{heafield2011kenlm,
  title={{KenLM}: Faster and smaller language model queries},
  author={Heafield, Kenneth},
  booktitle={Proceedings of the sixth workshop on statistical machine translation},
  pages={187--197},
  year={2011}
}

@inproceedings{pratap2019wav2letter++,
  title={Wav2letter++: A fast open-source speech recognition system},
  author={Pratap, Vineel and Hannun, Awni and Xu, Qiantong and Cai, Jeff and Kahn, Jacob and Synnaeve, Gabriel and Liptchinsky, Vitaliy and Collobert, Ronan},
  booktitle={Proc. ICASSP},
  pages={6460--6464},
  year={2019},
}

@article{mohamed2022self,
  title={Self-supervised speech representation learning: A review},
  author={Mohamed, Abdelrahman and Lee, Hung-yi and Borgholt, Lasse and Havtorn, Jakob D and Edin, Joakim and Igel, Christian and Kirchhoff, Katrin and Li, Shang-Wen and Livescu, Karen and Maal{\o}e, Lars and others},
  journal={JSTSP},
  year={2022},
  publisher={IEEE}
}

@inproceedings{chiu2022self,
  title={Self-supervised learning with random-projection quantizer for speech recognition},
  author={Chiu, Chung-Cheng and Qin, James and Zhang, Yu and Yu, Jiahui and Wu, Yonghui},
  booktitle={Proc. PMLR},
  //booktitle={International Conference on Machine Learning},
  pages={3915--3924},
  year={2022},
  //organization={PMLR}
}

@article{zaiem2025speech,
  title={Speech self-supervised representations benchmarking: a case for larger probing heads},
  author={Zaiem, Salah and Kemiche, Youcef and Parcollet, Titouan and Essid, Slim and Ravanelli, Mirco},
  journal={Computer Speech \& Language},
  pages={101695},
  year={2025},
  publisher={Elsevier}
}

@inproceedings{pasad2021layer,
  title={Layer-wise analysis of a self-supervised speech representation model},
  author={Pasad, Ankita and Chou, Ju-Chieh and Livescu, Karen},
  booktitle={Proc. ASRU},
  pages={914--921},
  year={2021},
}

@incollection{hotelling1992relations,
  title={Relations between two sets of variates},
  author={Hotelling, Harold},
  booktitle={Breakthroughs in statistics: methodology and distribution},
  pages={162--190},
  year={1992},
  publisher={Springer}
}

@inproceedings{chung2021similarity,
  title={Similarity analysis of self-supervised speech representations},
  author={Chung, Yu-An and Belinkov, Yonatan and Glass, James},
  booktitle={Proc. ICASSP},
  pages={3040--3044},
  year={2021},
}

@inproceedings{schatz2013evaluating,
  title={Evaluating speech features with the minimal-pair ABX task: Analysis of the classical MFC/PLP pipeline},
  author={Schatz, Thomas and Peddinti, Vijayaditya and Bach, Francis and Jansen, Aren and Hermansky, Hynek and Dupoux, Emmanuel},
  booktitle={Proc. INTERSPEECH},
  pages={1--5},
  year={2013}
}

@article{dunbar2021zero,
  title={The zero resource speech challenge 2021: Spoken language modelling},
  author={Dunbar, Ewan and Bernard, Mathieu and Hamilakis, Nicolas and Nguyen, Tu Anh and De Seyssel, Maureen and Roz{\'e}, Patricia and Rivi{\`e}re, Morgane and Kharitonov, Eugene and Dupoux, Emmanuel},
  journal={arXiv preprint arXiv:2104.14700},
  year={2021}
}

@inproceedings{pasad2023comparative,
  title={Comparative layer-wise analysis of self-supervised speech models},
  author={Pasad, Ankita and Shi, Bowen and Livescu, Karen},
  booktitle={Proc. ICASSP},
  pages={1--5},
  year={2023},
}

@inproceedings{pwcca,
  title={Insights on representational similarity in neural networks with canonical correlation},
  author={Morcos, A and Raghu, M. and S. Bengio},
  booktitle={Proc. NeurIPS},
  year={2018}
}

@article{hadi2023survey,
  title={A survey on large language models: Applications, challenges, limitations, and practical usage},
  author={Hadi, Muhammad Usman and Qureshi, Rizwan and Shah, Abbas and Irfan, Muhammad and Zafar, Anas and Shaikh, Muhammad Bilal and Akhtar, Naveed and Wu, Jia and Mirjalili, Seyedali and others},
  journal={Authorea Preprints},
  year={2023},
  publisher={Authorea}
}

@article{brown2020language,
  title={Language models are few-shot learners},
  author={Brown, Tom and Mann, Benjamin and Ryder, Nick and Subbiah, Melanie and Kaplan, Jared D and Dhariwal, Prafulla and Neelakantan, Arvind and Shyam, Pranav and Sastry, Girish and Askell, Amanda and others},
  journal={Advances in neural information processing systems},
  volume={33},
  pages={1877--1901},
  year={2020}
}

@article{li2023starcoder,
  title={{StarCoder}: may the source be with you!},
  author={Li, Raymond and Allal, Loubna Ben and Zi, Yangtian and Muennighoff, Niklas and Kocetkov, Denis and Mou, Chenghao and Marone, Marc and Akiki, Christopher and Li, Jia and Chim, Jenny and others},
  journal={arXiv preprint arXiv:2305.06161},
  year={2023}
}

@article{ling2017program,
  title={Program induction by rationale generation: Learning to solve and explain algebraic word problems},
  author={Ling, Wang and Yogatama, Dani and Dyer, Chris and Blunsom, Phil},
  journal={arXiv preprint arXiv:1705.04146},
  year={2017}
}

@article{vashishth2024stab,
  title={{STAB}: speech tokenizer assessment benchmark},
  author={Vashishth, Shikhar and Singh, Harman and Bharadwaj, Shikhar and Ganapathy, Sriram and Asawaroengchai, Chulayuth and Audhkhasi, Kartik and Rosenberg, Andrew and Bapna, Ankur and Ramabhadran, Bhuvana},
  journal={arXiv preprint arXiv:2409.02384},
  year={2024}
}
}
%\begin{thebibliography}{00}

%\bibitem{b1} G. Eason, B. Noble, and I. N. Sneddon, ``On certain integrals of Lipschitz-Hankel type involving products of Bessel functions,'' Phil. Trans. Roy. Soc. London, vol. A247, pp. 529--551, April 1955.
%\bibitem{b2} J. Clerk Maxwell, A Treatise on Electricity and Magnetism, 3rd ed., vol. 2. Oxford: Clarendon, 1892, pp.68--73.
%\bibitem{b3} I. S. Jacobs and C. P. Bean, ``Fine particles, thin films and exchange anisotropy,'' in Magnetism, vol. III, G. T. Rado and H. Suhl, Eds. New York: Academic, 1963, pp. 271--350.
%\end{thebibliography}
\vspace{12pt}

%\begin{thebibliography}{00}
%\input{conference_101719.bbl}
%\end{thebibliography}

\end{document}